\documentclass[prb,twocolumn,tightenlines,floatfix,showpacs,superscriptaddress]{revtex4}

\usepackage{graphicx}

\begin{document}

\bibliographystyle{prsty} 

\title{Nuclear Magnetic Resonance in a Ferromagnet-Semiconductor Heterostructure}

\author{J. Strand}
\affiliation{School of Physics and Astronomy}
\author{B. D. Schultz}
\affiliation{Department of Chemical Engineering and Materials
Science, \\University of Minnesota, Minneapolis, MN 55455}
\author{A. F. Isakovic}
\affiliation{School of Physics and Astronomy}
\author{X. Lou}
\affiliation{School of Physics and Astronomy}
\author{C. J. Palmstr{\o}m}
\affiliation{Department of Chemical Engineering and Materials
Science, \\University of Minnesota, Minneapolis, MN 55455}
\author{P. A. Crowell}\email{crowell@physics.umn.edu}
\affiliation{School of Physics and Astronomy}

\begin{abstract}
We report the observation of nuclear magnetic resonance (NMR) in a
ferromagnet-semiconductor heterostructure in the presence of a
spin-polarized current. Spin-polarized electrons injected from a
metallic ferromagnet generate a large nuclear spin population in a
GaAs quantum well by dynamic polarization. The characteristic time
for the polarization process is approximately 20 sec, and  the
nuclear polarization can persist for several minutes after the
current is turned off.  Resonant depolarization is observed in the
presence of an AC magnetic field or when the injection current is
modulated at the NMR frequency.

\end{abstract}


\maketitle


A detailed understanding of the hyperfine interaction is a
prerequisite for the coherent control of electron and nuclear spin
systems in semiconductor heterostructures.  Electrical spin
injection\cite{Molenkamp_Ohno_Ploog_Jonker_IMEC} provides a simple
means of introducing spin-polarized carriers into a semiconductor,
but it has proven more difficult to achieve the degree of tunability
over the combined electron and nuclear spin systems demonstrated
in optical pumping
experiments.\cite{Kikkawa:Awschalom:allOptNMR,Salis:OptNMR-PRL,
Epstein:Awschalom:FPP_in_n-GaAs,Stephens:Awschalom:Imaging}  For example,
Poggio {\it et al.}\cite {Poggio:Awschalom:gated_NMR}
have recently shown that it is possible to electrically
control the spatial profile of optically-pumped
nuclear polarization within a single quantum well.
One of
the advantages of optical pumping is the efficiency of dynamic
nuclear polarization (DNP), in which a non-equilibrium nuclear
polarization is generated by the spin relaxation of
electrons.\cite{Overhauser,Lampel,Paget:GaAsNuclearSpinCoupling,NucMag:Abragam}
In principle, the DNP process can also be driven by a
spin-polarized current.\cite{Johnson:ElectricalDNP-proposal} This
was inferred in recent experiments on Fe/Al$_x$Ga$_{1-x}$As
heterosructures,\cite{Strand:Crowell:electrical-DNP} although an
explicit dynamical signature of DNP was not demonstrated.

In this letter we report the direct observation of nuclear
magnetic resonance (NMR) in a GaAs quantum well (QW) in the
presence of a spin-polarized current.  We demonstrate that DNP can
be partially supressed by resonant depolarization of the nuclear
spins.  The depolarization can be achieved either by the
traditional approach of applying a resonant AC magnetic field or
by modulating the spin-polarized current at the NMR frequency. The
time dependence of the DNP process is measured, indicating a
characteristic build-up time of $\approx 20$~sec and persistence
of nuclear polarization up to eight minutes after the
spin-polarized current has been turned off.

\begin{figure}
    \includegraphics{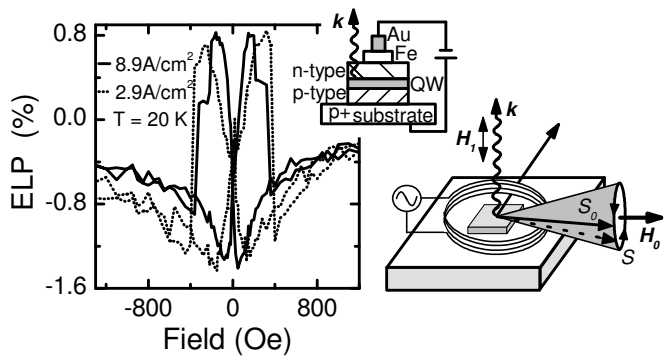}
    \caption{Electroluminescence polarization as a function of the applied
    field $H_0$ for two current densities.  Higher current corresponds to
    larger nuclear field. Center inset: cross-sectional schematic of the device. Right
    inset: schematic  of the experimental setup. The device is at the center
    of a coil which generates an AC magnetic field ${\bf H}_1$ out of the plane,
    parallel to the direction of light propagation $\bf k$.
    ${\bf H}_0$ is in the device plane.  The electron spin $\bf S$
    precesses about ${\bf H}_0$ after injection from Fe into the GaAs. The
    angle between the initial electron spin ${\bf S}_0$ and ${\bf H}_0$ is determined by
    the anisotropy of the Fe film.}
    \label{fig:cartoons}
\end{figure}

The heterostructure
is grown by molecular beam epitaxy on a p$^+$ GaAs (100)
substrate and consists of
p-Al$_{0.1}$Ga$_{0.9}$As/QW/n-Al$_{0.1}$Ga$_{0.9}$As/Fe/Al. The QW
comprises 100~\AA~GaAs between intrinsic Al$_{0.1}$Ga$_{0.9}$As barriers,
and the Fe film is 50~\AA~thick.
The
metal layers are grown {\it in situ} at $\approx 0^\circ$~C, and a
$3\times 10^{13}$~atoms/cm$^{2}$ $\delta$-doped layer of Si is
grown 25~{\AA} from the Fe/Al$_x$Ga$_{1-x}$As interface to create a thin Schottky
tunnel barrier.
The devices are fabricated into bars with an active region 80~$\mu$m wide (see schematic drawing in
Fig.~\ref{fig:cartoons}), and the experiments are carried out in a
magneto-optical cryostat at temperatures ranging from 2~K to 40~K.

The measurements are performed in the Voigt geometry with the
applied field ${\bf H}_0$ in the plane of the QW.
Electroluminescence (EL) is collected along the growth direction,
indicated by ${\bf k}$ in Fig.~\ref{fig:cartoons}, which coincides
with the direction of the AC magnetic field ${\bf H}_1$ generated
by a four-turn coil placed on top of the sample. The EL is due to
the recombination of electrons tunneling from the Fe layer with
heavy holes flowing from the substrate.  The  circular
polarization of the EL, referred to as the ELP signal, is equal to
the component of electron spin polarization along  $\bf {k}$ at
the time of recombination.

Typical ELP data as a function of applied field are shown in
Fig.~\ref{fig:cartoons}. All of the measurements under discussion
here were conducted with ${\bf H}_0$ nearly parallel to the
$[01{\overline 1}]$ direction, which corresponds to a magnetic
hard axis.  As $H_0$ decreases below 500~Oe, the magnetization
${\bf M}$ rotates away from $[01\overline 1]$ towards the $[011]$
(easy) direction.  The non-zero angle between the injected spin
${\bf S}_0 \parallel {\bf M}$ and ${\bf H}_0$ leads to precession
of the spin in the QW, resulting in a steady-state component of
${\bf S}$ along ${\bf k}$ and hence the observed ELP signal. The
ELP hysteresis loops shown in Fig.~\ref{fig:cartoons} can be
modeled extremely well provided that the total field ${\bf B} =
{\bf H}_0 +{\bf B}_N$, where ${\bf B}_N$ is assumed to be the
hyperfine field due to nuclei polarized by DNP. As discussed in
Ref.~\onlinecite{Strand:Crowell:electrical-DNP}, the magnitude of
${\bf B}_N$ depends on the injection current density and the angle
$\theta$ between ${\bf S}_0$ and ${\bf H}_0$ and reaches several
kilogauss for current densities $\sim 10$~A/cm$^2$ and $\theta <
\pi/2$.

\begin{figure}
    \includegraphics{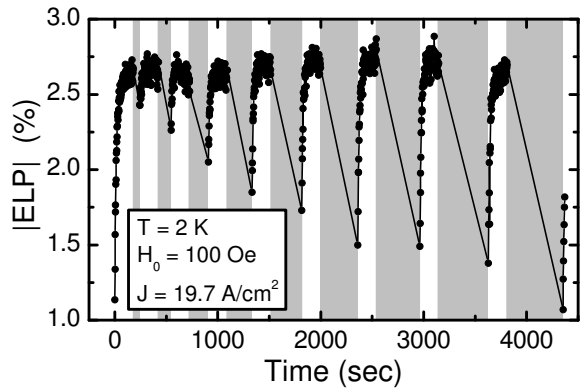}
    \caption{Laboratory time dependence of the electroluminescence
    polarization (ELP) signal.  The device current is left off for
    ten minutes prior to $t = 0$, and the current is then switched between
    on (three minutes) and off (gray regions), with the $t_{\rm off}$ increasing
    from 1 minute to 9 minutes.}
    \label{fig:time}
\end{figure}

The large nuclear magnetic field is due to the long
spin-relaxation time for nuclei and the large difference between
the electron and nuclear magnetic moments.\cite{NucMag:Abragam}
Since the nuclear spin-lattice relaxation time can be many
minutes,\cite{OO,Kikkawa:Awschalom:allOptNMR} we expect the field
${\bf B}_N$ to build up on laboratory time scales when the current
is turned on and then decay slowly when the current is turned off.
Figure~\ref{fig:time} shows the ELP signal as the device current
is switched on and off for variable amounts of time. In this case,
the current is off for ten minutes prior to the start of data
collection at $t = 0$. The LED is then turned on and the ELP
signal is measured for three minutes.  The increase in the ELP
signal is due to the increase in total internal magnetic field and
the consequent increase in the angle of precession of the injected
spin polarized electrons. To probe the lifetime of the nuclear
polarization in the absence of continuous pumping, the current is
turned off for a time $t_{\rm off}$ and then on again for three
minutes.   Figure~\ref{fig:time} shows a series of these
measurements with $t_{\rm off}$ varying from 1 min to 9 min.  The
first data point taken after each ``off'' period indicates the
degree of nuclear polarization that has been retained.  By
repeating this procedure with increasing $t_{\rm off}$, we observe
a decay time of approximately 5 min. Although both the build-up
and decay times for the ELP signal are long, it is difficult to
make a direct quantitative mapping between the magnitude of the
ELP signal and the actual nuclear polarization.   In particular,
the relationship between the ELP signal and $B_N$ is very
non-linear. The ELP signal increases rapidly at small nuclear
fields and then saturates,\cite{Strand:Crowell:electrical-DNP} and
this is responsible in part for the large difference between the build-up
and decay times for the ELP signal.

\begin{figure}
    \includegraphics{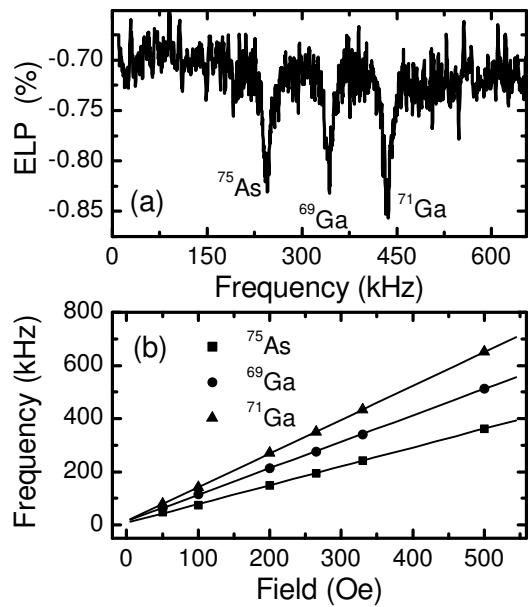}
    \caption{(a) Frequency-swept NMR depolarization features at T~=~20~K,
    $H_0 = 330$~Oe, and $H_1 \approx 0.2$~Oe, corresponding to the three
    isotopes present in the GaAs QW. (b) Depolarization peak frequency as
    a function of $H_0$ at 20~K and $H_1 \approx 0.2$~Oe. The solid lines
    are linear fits.}
    \label{fig:peaks}
\end{figure}

Although the nuclear polarization can relax slowly through coupling to the lattice,
the most effective way of destroying DNP is to apply an
oscillating magnetic field ${\bf H}_1$ at the NMR frequency along a direction perpendicular
to ${\bf H}_0$.  The transverse field couples directly to the nuclear spin system, leading to
 depolarization of the nuclei that are in resonance.  $H_1$ is generated by a 1~cm
diameter, four turn coil driven by the sinusoidal output
of a function generator.
By sweeping the frequency of $H_1$ at fixed $H_0$,
depolarization signatures for each of the three isotopes in the
GaAs QW [$^{75}$As ($\gamma = 0.731$~kHz/Oe), $^{69}$Ga ($\gamma =
1.025$~kHz/Oe), and $^{71}$Ga ($\gamma = 1.302$~kHz/Oe)]\cite {CRCgamma} are
observed, as seen in Fig.~\ref{fig:peaks}(a) for $H_0 = 330$~Oe.
 In each
case, the linewidth of the depolarization signature is 10 - 15 kHz, with no systematic dependence
on magnetic field.  The linewidth at these fields ($< 500$~Oe) is therefore not
limited by inhomogeneous dephasing.

Frequency sweeps were conducted at several different magnetic
fields. The peaks of the depolarization features for all three
isotopes are plotted versus  $H_0$ in Fig.~\ref{fig:peaks}(b).
Linear fits of $\nu = \gamma H_0$ were obtained for each isotope,
yielding gyromagnetic ratios $\gamma_{^{75}\rm As} =
0.71\pm0.1$~kHz/Oe, $\gamma_{^{69}\rm Ga} = 1.00\pm0.04$~kHz/Oe,
and $\gamma_{^{71}\rm Ga}= 1.27\pm0.04$~kHz/Oe, in agreement with
the accepted values.  The deviations from the fits are smaller
than the limit set by trapped flux in the magnet ($\approx
10$~Oe).

The sign of the depolarization features warrants comment, since
the dips in Fig.~\ref{fig:peaks}(a) correspond to an increase in
the magnitude of the ELP signal, which is negative in this field
range. We have confirmed that the features in Fig.~\ref
{fig:peaks}(a) correspond to a {\it decrease} in the nuclear
polarization by comparing with the observed dependence of the ELP
signal on current density.\cite{Strand:Crowell:electrical-DNP} The
reason for this non-intuitive result is that for sufficiently
large nuclear fields, the average angle through which the
electrons precess during their lifetime becomes larger than
$\pi/2$.  An increasing nuclear field therefore results in a
decrease in the magnitude of the ELP signal because the electrons
are actually precessing further away from the measurement axis.
Although a complete quenching of $B_N$ would result in nearly
total suppression of the ELP signal, only a single isotope is
depolarized at each resonance.  We have  confirmed that for low
$B_N$, corresponding to applied fields less than 60~Oe, resonant
depolarization results in a decrease in the ELP magnitude as
expected.

\begin{figure}
    \includegraphics{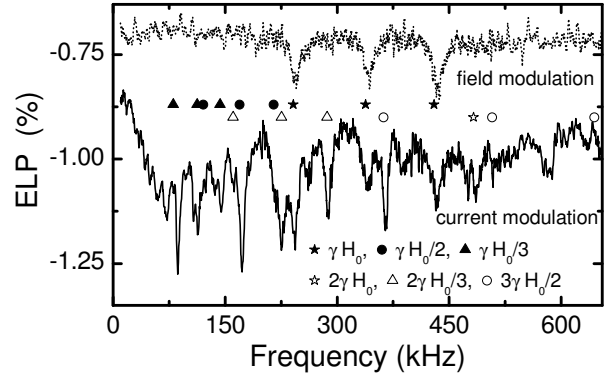}
    \caption{Frequency-swept NMR depolarization features at
T~=~20~K, and $H_0 = 330$~Oe corresponding to the three isotopes
present in the GaAs QW. The upper curve shows the ELP signal
measured with an AC magnetic field $H_1 \approx 0.2$~Oe. The lower
curve shows the ELP signal measured under alternating device
current ($H_1 = 0$~Oe). The filled symbols identify principal
nuclear transition frequencies and their sub-harmonics. The empty
symbols indicate harmonics of the primary resonances as well as
their sub-harmonics.}
    \label{fig:current}
\end{figure}

Since the nuclear polarization in these devices is generated by
the injected electrons, an alternative approach to depolarization
is resonant modulation of the device current.  In this case, the
coil is not used ($H_1 = 0$), and the DC bias voltage is set above
the threshold for light emission.   An AC voltage is then added so
that the average bias voltage corresponds approximately to the
value used in the measurements discussed above.
 In this case, the bias is
being used only to modulate the injection current.  Given the
square well design, we cannot control the relative position of the
nuclear and electronic spins as achieved recently in parabolic
quantum wells.\cite {Poggio:Awschalom:gated_NMR}
Fig.~\ref{fig:current} shows the ELP signal as a function of bias
voltage frequency for $H_0 = 330$~Oe, $V_{DC} = 2.1$~V, and
$V_{AC}=0.5$~V$_{PP}$. The principal resonances are clearly
visible, as can be seen by comparing the data taken under
alternating device current (lower curve) with the data taken under
AC magnetic field (upper curve). In the case of alternating device
current, however, there are many other peaks in the data, and
frequency sweeps at different fields indicate that these also
correspond to nuclear resonances. Some of the additional peaks can
be mapped to sub-harmonics of the primary resonances, such as $\nu
= \gamma H_0 /3$.  Other peaks occur at either harmonics of the
primary resonances, such as $\nu = 2 \gamma H_0$, or their
sub-harmonics, such as $\nu = 2\gamma H_0 /3$. Similar harmonic
and subharmonic resonances have also been observed in quantum well
systems with quadrupolar interactions in which a strong electric
field gradient is modulated with a gate
voltage.\cite{Poggio:Awschalom:gated_NMR}

Sub-harmonics of the primary resonances are never observed with an
AC magnetic field, and we attribute them to the non-sinusoidal
time dependence of the injection current  as the bias voltage is
modulated.  This is a consequence of the diode structure of the
device. The origins of harmonics of the fundamental resonances are
less clear.  These peaks are also observed in AC magnetic field
experiments at high amplitudes ($H_1 > 0.4$~Oe).   The observation
of the $\Delta m = \pm 2$ peaks in the AC magnetic field
experiments at high amplitudes indicates that in this experiment
they originate from some intrinsic non-linearity. Dipole-dipole
coupling can lead to both $\Delta m = \pm 2$ and $\Delta m =\pm 3$
transitions,\cite{Kalevich:Fleisher:2spin_resonance_GaAs} and may
play a more significant role because of the large nuclear
polarization in these experiments.

The spectroscopic measurements presented here unambiguously demonstrate dynamic
nuclear polarization by
electrical spin injection.   The ability to modulate the nuclear
polarization with a spin-polarized current provides a unique means of controlling the coupled
electron-nuclear spin system.
This capability, in combination with the large nuclear polarization
achieved at low fields, provides a new approach to NMR spectroscopy of heterostructures.

We thank S. A. Crooker and M. S. Conradi for helpful discussions.
This work was supported by ONR, the DARPA/ONR SPINS program, and the
University of Minnesota MRSEC (NSF DMR-0212032). A.F.I. and B.D.S.
thank 3M for support.


\begin{thebibliography}{10}

\bibitem{Molenkamp_Ohno_Ploog_Jonker_IMEC}
R. Fiederling {\it et~al.}, Nature {\bf 402},  787  (1999); Y.
Ohno {\it et~al.}, Nature {\bf 402},  790  (1999); H.~J. Zhu {\it
et~al.}, Phys. Rev. Lett. {\bf 87},  016601  (2001); A.~T.
Hanbicki {\it et~al.}, Appl. Phys. Lett. {\bf 80},  1240 (2002);
V.~F. Motsnyi {\it et~al.}, Appl. Phys. Lett. {\bf 81}, 265
(2002).

\bibitem{Kikkawa:Awschalom:allOptNMR}
J.~M.~Kikkawa and D.~D.~Awschalom, Science {\bf 287}, 473.

\bibitem{Salis:OptNMR-PRL}
G. Salis {\it et~al.}, Phys. Rev. Lett. {\bf 86},  2677
   (2001).

\bibitem{Epstein:Awschalom:FPP_in_n-GaAs}
R.~J. Epstein {\it et~al.}, Phys. Rev. B {\bf 65},  121202 (2002);
R.~J. Epstein {\it et~al.}, Phys. Rev. B {\bf 68},  041305 (2003);

\bibitem{Stephens:Awschalom:Imaging}
J.~Stephens {\it et al.}, cond-mat/0303448.

\bibitem{Poggio:Awschalom:gated_NMR}
M.~Poggio {\it et~al.}, cond-mat/0306012.

\bibitem{Overhauser}
A. W. Overhauser, Phys. Rev. {\bf 92}, 411 (1953).

\bibitem{Lampel}
G. Lampel, Phys. Rev. Lett. {\bf 20}, 491 (1968).

\bibitem{Paget:GaAsNuclearSpinCoupling}
D. Paget {\it et~al.}, Phys. Rev. B {\bf 15},
  5780  (1977).

\bibitem{NucMag:Abragam} A. Abragam, {\it Principles of Nuclear Magnetism}
(Oxford University Press, Oxford, 1961).


\bibitem{Johnson:ElectricalDNP-proposal}
M. Johnson, Appl. Phys. Lett. {\bf 77},  1680  (2000).


\bibitem{Strand:Crowell:electrical-DNP}
J.~Strand {\it et~al.}, Phys. Rev. Lett. {\bf 91}, 036602 (2003).

\bibitem{OO}
{\em Optical Orientation}, edited by F. Meier and B.~P.
Zakharchenya (North-Holland Physics Publishers, New York, 1984).

\bibitem{CRCgamma} {\em CRC Handbook of Chemistry and Physics, 76th ed.}, edited by D.~R.~Lide
(CRC Press, Boca Raton, 1995).

\bibitem{Kalevich:Fleisher:2spin_resonance_GaAs}
V.~K.~Kalevich {\it et~al.}, Sov. Phys. Solid State {\bf 24}, 1195
(1982).

\end{thebibliography}
\end{document}